\documentclass[showpacs,preprintnumbers,preprint,amsmath,amssymb, graphicx, graphics, epsfig]{revtex4}
\input{epsf}

\usepackage{epsfig}
\usepackage{dcolumn}
\usepackage{bm}

\begin{document}


\def\be{\begin{equation}}
\def\ee{\end{equation}}
\def\bdm{\begin{displaymath}}
\def\edm{\end{displaymath}}
\def\erfc{\hbox{erfc }}
\def\vpa{v_{\parallel }}
\def\vper{v_{\perp }}
\def\Omm{\Omega }
\def\ppa{p_{\parallel }}
\def\pper{p_{\perp }}
\def\ppv{\vec{p}}
\def\kkv{\vec{k}}
\def\omm{\omega}
\def\krz{{\bf \times }}
\def\dele{\delta \vec{E}(\vec{k},\omega )}
\def\delb{\delta \vec{B}(\vec{k},\omega )}
\def\Ab{\sum_a\omm _{p,a}^2(m_ac)^3}
\def\erf{\rm {erf }}

\title{Fast magnetization in counterstreaming plasmas with temperature anisotropies}

\author{M. Lazar$^1$}
\email{mlazar@tp4.rub.de}
\affiliation{$^1$ Institut f\"ur Theoretische Physik, Lehrstuhl IV:
Weltraum- und Astrophysik, Ruhr-Universit\"at Bochum, D-44780 Bochum, Germany}
\altaffiliation[Also at ]{the Department of Physics and Engineering Physics, University of Saskatchewan, 116 Science Place, Saskatoon, Saskatchewan S7N 5E2, Canada }

\date{\today}

\begin{abstract}
Counterstreaming plasmas exhibits an electromagnetic unstable 
mode of filamentation type, which is responsible for the
magnetization of plasma system. It is shown 
that filamentation instability becomes significantly faster
when plasma is hotter in the streaming direction. 
This is relevant for astrophysical sources, where 
strong magnetic fields are expected to exist and
explain the nothermal emission observed.
\end{abstract}

\maketitle

\section{Introduction}

Presently, there is an increasing interest for a correct 
understanding of purely growing electromagnetic instabilities driven 
by a velocity anisotropy of plasma particles.
\cite{s04,s05,b06,lss06,zm07,y07}.
These instabilities will release the excess of 
perpendicular free energy stored in the particle velocity 
anisotropy, whether it is a temperature anisotropy \cite{w59} 
or counterstreaming plasmas \cite{f59}. A substantial fraction of the
kinetic energy of plasma particles is transformed by the instability
and contribute to the amplification of magnetic energy \cite{c98}.

Counterstreaming plasma structures are present in 
laboratory experiments and astrophysical systems,
and they are widely investigated either to prevent 
the unstable modes arising in beam-plasma 
experiments \cite{s02,bd06}, or to prove the existence of 
large scale magnetic fields in astrophysical objects 
\cite{ml99,sh03,ss03,oh03}.

The spontaneous magnetization of counterstreaming plasmas 
is associated with filamentation instability \cite{f59}, 
which is described as a counterstreaming based electromagnetic 
mode purely growing in time and propagating perpendicular 
to the streams. 
Here we show that the magnetization process associated 
with filamentation instability can be markedly faster
in counterstreaming plasmas with temperature anisotropies,
namely, when plasma is hotter along streaming direction.
This fact can be decisive for the potential role of
filamentation instability in producing strong quasistatic
magnetic fields in astrophysical plasmas \cite{c98,ml99}. 
As long the maximum growth rate of filamentation mode 
is much larger than the growth rates of all other plasma
instabilities (e.g., two-stream electrostatic instability), 
then the filamentation instability will be fastest, and 
it will be the primary mechanism for the relaxation of the initial 
counterstreaming configuration \cite{ss03}.

\section{Counterstreaming plasmas with temperature anisotropies}

The oscillatory properties of a plasma are determined by the 
dielectric tensor, $(\epsilon)$, and using the linearization of 
Vlasov-Maxwell equations we simply get \cite{kmq68}

\be
(\epsilon) ={\rm \bf I}+ \sum_a {\omm_{p,a}^2 \over \omega^2 }
\left[\int_{-\infty}^{+ \infty} \, d{\bf v} \; {\bf v}
\, {\partial f_{a,0} \over \partial {\bf v}} +
\int_{-\infty}^{+ \infty} \, d{\bf v} \,
{1 \over \omm - {\bf k}\cdot {\bf v}} \; ({\bf k}\cdot
{\partial f_{a,0} \over \partial {\bf v}}) \; {\bf v}{\bf v} \right]. \label{e1}
\ee
where the unperturbed velocity distribution function $f_{a0}({\bf v})$ 
of the particles of sort $a$ is normalized by $\int \, d{\bf v} \, f_{a0}({\bf v}) = 1$, 
$\omm_{p,a} = (4 \pi n_a e^2 /m_a)^{1/2}$ is the plasma frequency,
{\bf I} is the unity tensor, and $\omega$ and $k$ are respectively, the frequency and
the wave number of plasma modes.

\begin{figure}[h] \centering
    \includegraphics[width=60mm, height=45mm]{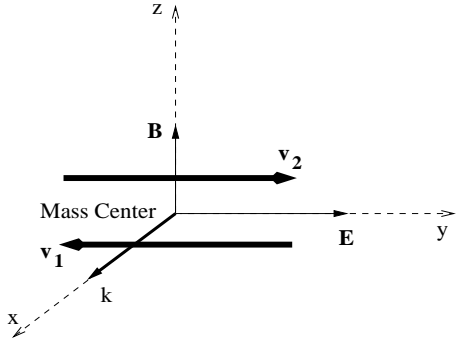}
\caption{Two counterstreaming plasmas and the electromagnetic 
filamentation mode with fields ${\bf E}$ and ${\bf B}$, 
and the wavevector ${\bf k}$ perpendicular to the streams.} \label{fig1} 
\end{figure}

In Figure \ref{fig1} we fix the orientation for the
counterstreams and the filamentation 
mode: the electric field, ${\bf E} \parallel {\bf v_{1,2}}$, 
is along the streaming direction, and the wave vector, 
${\bf k} \perp {\bf v_{1,2}}$, is perpendicular to the streams.
This electromagnetic mode will be solution of the following dispersion 
equation
\be
{k^2c^2 \over \omega^2} = \epsilon_{yy}. \label{e2}
\ee
with $\epsilon_{yy}$ given by (\ref{e1}), and $c$ 
is the speed of light in vacuum.

For the sake of simplicity, we assume that the counterstreams are symmetric 
with equal intensities, and moving with the same velocities, $|v_1| = v_2 = v_0$.
Moreover, the effect of ions is minimized to a positive background, and the electron 
plasma counterstreams are assumed to be homogeneous, and charge and current neutralized.

\subsection{Moderate thermal effects}

We first suppose that only one of the counterstreaming plasmas exibits 
a temperature anisotropy of a bi-Maxwellian type, i.e. with two 
characteristic thermal velocities, 
$v_{\rm th,x}= v_{\rm th,z}=v_{\rm th}< v_{\rm th,y}$, and the other one is monochromatic (cold)

\be
f_0 (v_x, v_y, v_z) = {1 \over 2 \pi^{3/2} v_{\rm th}^2 v_{{\rm th},y}} 
 e^{-{v_x^2+v_z^2 \over v_{\rm th}^2}} e^{-{(v_y+v_0)^2 \over v_{{\rm th},y}^2}}
 + {1\over 2} \delta(v_x) \delta(v_y - v_0)\delta(v_z). \label{e3}
\ee
We substitute (\ref{e3}) in (\ref{e1}), and then for dispersion relation (\ref{e2}) we find

\be
{k^2c^2 \over \omega^2} = 1 -{\omega_{pe}^2 \over 2\omega^2} \left(1+ {k^2 v_0^2 \over \omega^2}\right)
- {\omega_{pe}^2 \over 2\omega^2} \left\{1+ \left[{1\over 2}(A+1) 
+ \left({ \beta_0\over \beta_{\rm th}}\right)^2 \right] Z\,'\left({\omega \over k v_{\rm th}} \right) \right\},\label{e4}
\ee
where $\beta_{\rm th} = v_{\rm th}/c$, $\beta_0 = v_0/c$, 
and $Z \, '(f)$ is the derivative of the well-known plasma dispersion function \cite{fc61}

\be
Z(f) = \pi^{-1/2} \int_{-\infty}^\infty dx \, {\exp(-x^2) \over x-f}, \; \;\;
f = {\omega \over k v_{\rm th}}. \label{e5}
\ee

We assume that plasma is hotter in the streaming direction,
$v_{{\rm th},y} > v_{\rm th}$, which defines here a positive
temperature anisotropy $A_1 = A=(v_{{\rm th},y} /v_{\rm th})^2 -1 >0$. 
In this case the instability is driven equally by the counterstreaming motion of plasma
and by the temperature anisotropy of plasma particles, and therefore we should
achieve an enhancing effect for the growth rates of filamentation mode.

\begin{figure}[h] \centering
    \includegraphics[width=65mm, height=50mm]{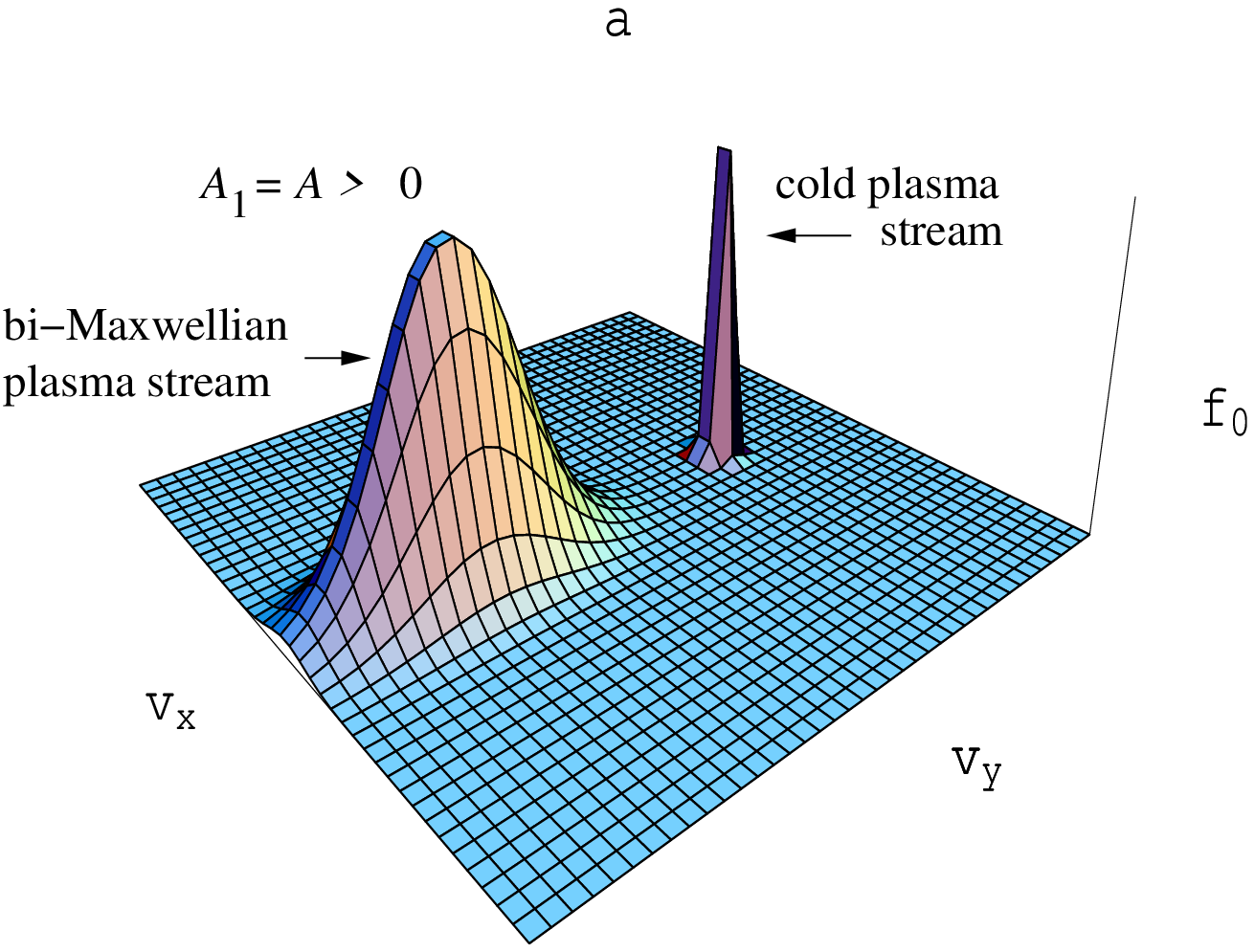} \;\;\;\;\;\;\;\;
    \includegraphics[width=55mm, height=55mm]{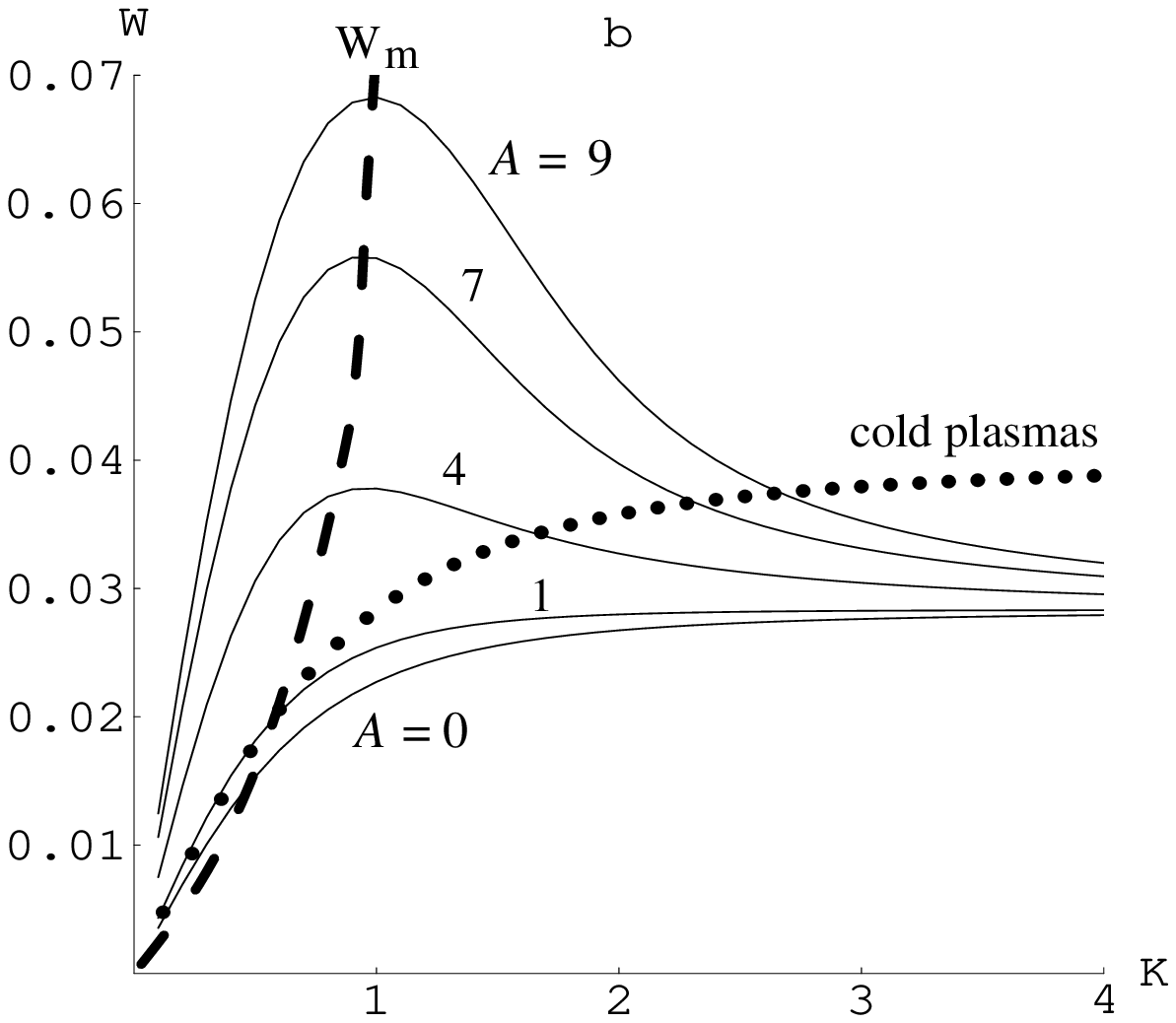}
\caption{(a) Schematic representation for the 
distribution functions of two counterstreaming plasmas: one is assuming cold,
and the other one with a two--temperatures anisotropy; 
(b) The growth rates of filamentation mode (solid lines) 
for $ \beta_{\rm th}= 0.1$, $\beta_0 = 0.04$, and 
different values of thermal anisotropy: $A =$9, 7, 4, 1, 0. 
The growth rates are also plotted for two cold counterstreaming plasmas (dotted 
line). The dashed line represents the {\it locus} of the maximum growth rates.
The coordinates are scaled as $W = \Omega /\omega_{pe}$ and 
$K = kc/\omega_{pe}$.} \label{fig2} 
\end{figure}

The aperiodic solutions of (\ref{e4}), $\Im (\omega)
= \Omega =\Omega (k)$ ($\Re (\omega) = 0$), are plotted 
in Figure \ref{fig2} (b)
with solid lines for different values of the positive anisotropy $A=0, 1, 4, 7, 9$, 
and with dotted line when both counterstreaming plasmas are cold. 
In this case, thermal effects are sufficiently low and 
the so-called "cold anisotropy" determined by the the 
relative motion of plasmas is still dominant leading to 
a mixture of the features from both filamentation and Weibel modes.
Therefore, for small wavenumbers the aperiodic solutions are 
characterized by a maximum, $\Omega_m = \Omega_m(k_m)$ 
and the corresponding wavenumber $k_m$ (Weibel regime), and 
for very large wavenumbers, $k \to \infty$, they approach an 
asymptotic value, $\Omega_{\rm a} = \Omega (k \to \infty)$ 
(filamentation regime). 
The last one is simply found as being lower than 
the asymptotic value obtained for the filamentation growth rate when the
both counterstreaming plasmas are cold, $\Omega_{\rm a}^{\rm cold}$,

\be
\Omega_{\rm a} = {\sqrt{2}\over 2} {\omega_{pe} v_0 \over c}
< \Omega_{\rm a}^{\rm cold} = {\omega_{pe} v_0 \over c}. \label{e6}
\ee

Knowing the maximum values for the growth rates of filamentation 
mode is essential for a correct evaluation of the magnetization 
process in laboratory or astrophysical applications \cite{ml99,d72}.
If the filamentation mode has the largest maximum growth rate
comparing to the other plasma instabilities (e.g., two-stream electrostatic
instability), then it will be the fastest mechanism by which is released
the free energy stored into the initial motion of plasma counterstreams.

The aperiodic solutions plotted in Fig. \ref{fig2} (b), are sensitive to values 
close to unity for the arguments of plasma dispersion 
function in (\ref{e4})--(\ref{e5}). Therefore, we keep the 
accuracy of a general approach, showing that the maximum 
growth rate, $\Omega_m$, can be determined exactly numerically without any 
restriction to large or small arguments of $Z (f)$.
The maximum growth rate, $\Omega_m$, depends on the temperature anisotropy 
(see in Fig. \ref{fig2}) and it is given in (\ref{e4}) by the following condition

\be
{d\Omega \over dk} =0. \label{e7}
\ee
But the maximum value, $\Omega_m$, is solution of (\ref{e4}) as well, and eliminating
the anisotropy from equations (\ref{e4}) and (\ref{e7}) we simply find \cite{kmq68} 

\be
\left[1+{\imath \Omega_m \over k_m v_{\rm th}}
Z \left( {\imath \Omega_m \over k_m v_{\rm th}}\right)\right]
\left[M_1 (\Omega_m, k_m)\left({2\Omega_m^2 \over k_m^2 v_{\rm th}^2} +3\right)-4\right]
= M_1 (\Omega_m, k_m), \label{e8}
\ee
where
\be
M_1 (\Omega_m, k_m) = {2 k_m^2 c^2 \over \omega_{pe}^2} - {k_m^2v_0^2 \over \Omega_m^2} +2.
\label{e9}
\ee
This is an implicit form equation for the maximum growth rates, [$k_m$, $\Omega_m$],
which we plot with dashed line in Fig. \ref{fig2} (b). Intersections of the 
dashed line with dispersion curves given by (\ref{e4}) and plotted with solid lines 
in Fig. \ref{fig2} (b), will give us the maximum values corresponding to different 
temperature anisotropies, $\Omega_m = \Omega_m (k_m, A)$.

We remark that only the large wave-lengths will be affected by 
the temperature effects (the Weibel-like regime) where the amplitude is
growing significantly faster. At saturation, the growth rates can be
markedly larger than those obtained for isotropic counterstreams ($A=0$), 
or for cold plasmas. This enhancing effect is diminished at small wave-lenghts,
where the growing mode is not affected by the temperature anisotropy.

\subsection{Strong thermal effects}

\begin{figure}[h] \centering
    \includegraphics[width=65mm, height=50mm]{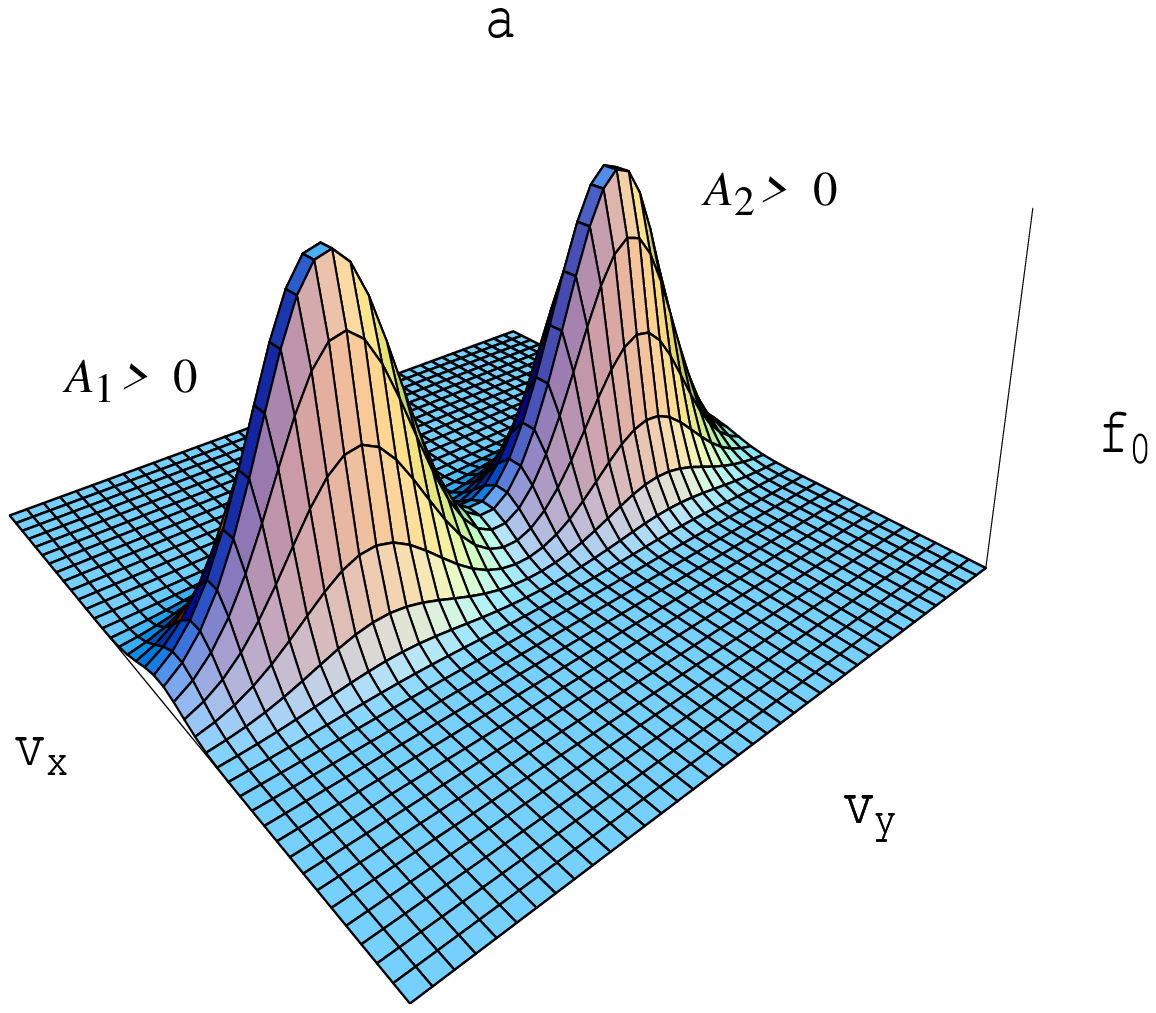}
    \includegraphics[width=65mm, height=50mm]{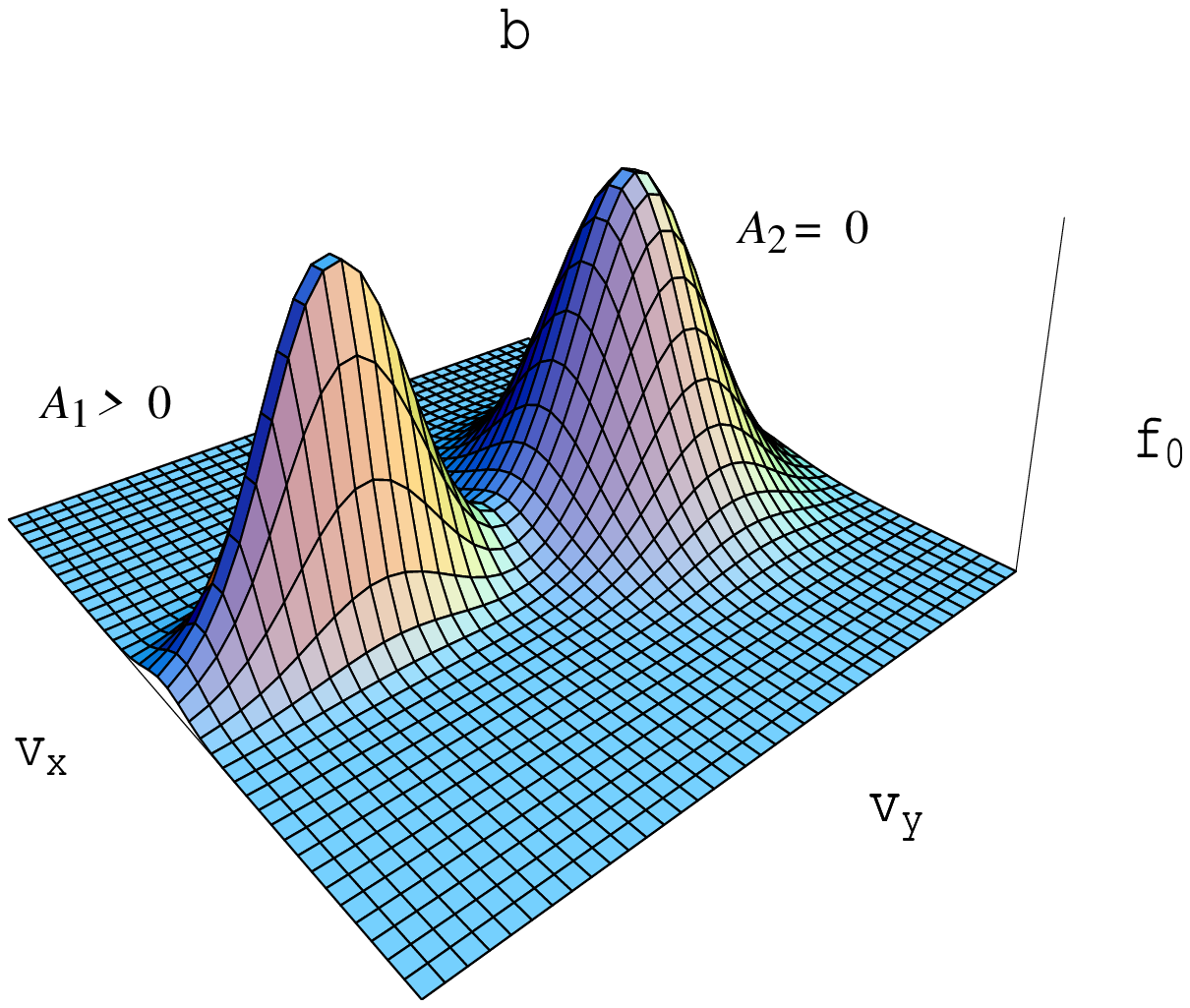}
    \includegraphics[width=60mm, height=55mm]{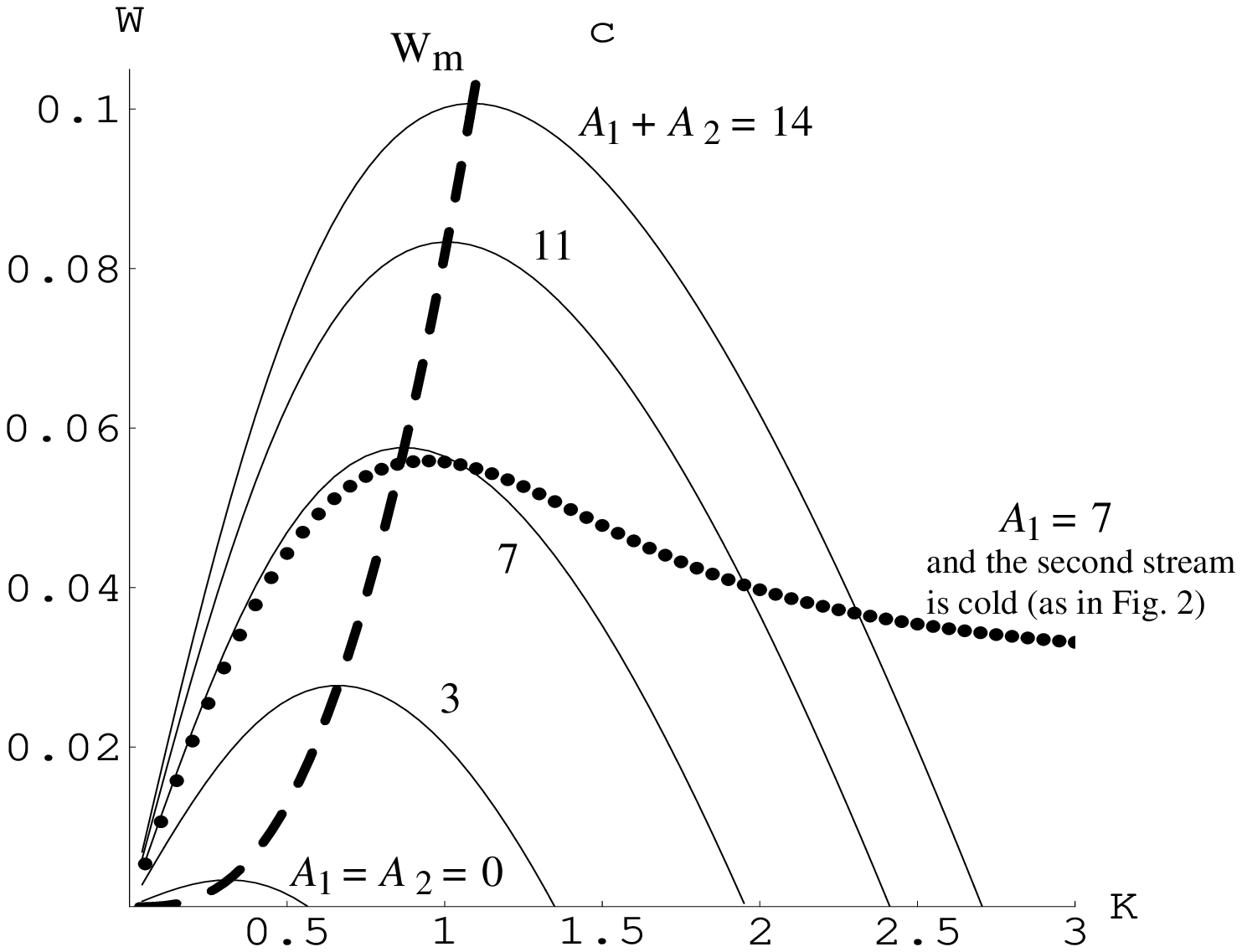}
\caption{(a) and (b) Schematic representation for the 
distribution functions of counterstreaming plasmas 
dominated by thermal effects. (c) With solid lines are shown
the aperiodic solutions of (\ref{e11}) for $\beta_{\rm th}= 0.1$, 
$\beta_0 = 0.04$ and different temperature anisotropies 
$A_1 + A_2=$14, 11, 7, 3, or $A_1 + A_2=0$. The dashed line 
represents the {\it locus} of the maximum growth rates. 
The coordinates are scaled as in Fig. \ref{fig2}.}\label{fig3} 
\end{figure}

We assume now that both plasma counterstreams are dominated by the thermal effects, 
with bi-Maxwellian distributions of positive anisotropies, 
$A_{1,2}= (v_{{\rm th},y 1,2} /v_{\rm th})^2 -1 \geqslant 0$.
In this case the counterstreams are modeled by the following
distribution function

\be
f_0 (v_x, v_y, v_z) = {1 \over 2 \pi^{3/2} v_{\rm th}^2 v_{{\rm th},y}} 
 e^{-{v_x^2+v_z^2 \over v_{\rm th}^2}}\left[ e^{-{(v_y+v_0)^2 \over v_{{\rm th},y,1}^2}}
 +e^{-{(v_y-v_0)^2 \over v_{{\rm th},y,2}^2}} \right]\;, \label{e10}
\ee
which is schematically presented in Fig. \ref{fig3} (a), and 
substituted in (\ref{e1}) yields the following dispersion relation

\be
{k^2c^2 \over \omega^2} = 1 - {\omega_{pe}^2 \over \omega^2} \left\{1+ {1\over 2}\left[1+ 
{1\over 2}(A_1+A_2) + 2 \left({ \beta_0\over \beta_{\rm th}}\right)^2\right] 
Z\,'\left({\omega \over k v_{\rm th}} \right) \right\}.\label{e11}
\ee
Due to the strong temperature effects the aperiodic solutions from (\ref{e11}), 
$\Omega(k) \geqslant 0$, appear to resemble the Weibel regime, 
see the solid line curves in Fig. \ref{fig3} (b), and the unstable mode does not 
exist for $k \geqslant k_c$, where the cutoff wave number $k_c$
is the nontrivial solution of (\ref{e8}) to the limit of $\Omega = 0$

\be
k_c (A_1, A_2)= {\omega_{pe} \over c} \left[{1\over 2}(A_1+A_2) + 2 \left({ \beta_0\over \beta_{\rm th}}\right)^2\right]^{1/2}.\label{e12}
\ee

For an exact evaluation of the maximum growth rate, [$k_m$, $\Omega_m$], we invoke again condition (\ref{e7})
and for the dispersion relation (\ref{e11}) we derive as before, an implicit form equation 

\be
\left[1+{\imath \Omega_m \over k_m v_{\rm th}}
Z \left( {\imath \Omega_m \over k_m v_{\rm th}}\right)\right]
\left[M_2 (\Omega_m, k_m)\left({2\Omega_m^2 \over k_m^2 v_{\rm th}^2} +1\right)+2\right]
= M_2 (\Omega_m, k_m), \label{e13}
\ee
with
\be
M_2 (\Omega_m, k_m) = {\omega_{pe}^2 \over k_m^2 c^2 } + { \Omega_m^2 \over k_m^2 c^2} +1.
\label{e14}
\ee
The maximum growth rates from (\ref{e13}) are plotted with dashed line 
in Fig. \ref{fig3} (c), and the intersections with dispersion curves 
given by (\ref{e11}) and plotted with solid lines, 
will give us the maximum values corresponding to different 
temperature anisotropies, $\Omega_m = \Omega_m (k_m, A)$.
In this case thermal effects limit the existence of 
the unstable mode to large wave-lengths, $k \leqslant k_c$, 
where $k_c$ is given by (\ref{e12}). This limit is removed, 
$k_c \to \infty$, only for $\beta_{\rm th} \to 0$.
However, the growth rates can be orders of magnitude
larger as long the positive temperature anisotropies are
sufficiently large.

\section{Conclusions}

We have investigated the purely growing electromagnetic mode
that arises in counterstreaming plasmas and propagates perpendicular
to the streaming direction. This instability is responsible
for the magnetization of such plasma systems. We have generalized 
the theoretical approach assuming that counterstreaming 
plasmas exhibit positive temperature anisotropies of a bi-Maxwellian type. 
In this case the instability is driven equally by the 
free energy stored in the counterstreaming motion of plasma
and the temperature anisotropy of plasma particles. 
Consequently, we found that the aperiodic modes can reach
maximum growth rates with order of magnitudes larger than
those calculated for cold plasmas or for counterstreaming plasmas
with an isotropic temperature distribution.
Thus, the filamentation mode is enhanced by the positive 
temperature anisotropies (i.e., when plasma is hotter in the streaming direction), 
and it can become faster than all the other plasma instabilities.
This effect improves the efficiency of magnetic field generation, 
and gives further support for the potential role of magnetic instabilities 
in the fast magnetization applications.

\acknowledgements
{This work was supported by the Alexander von Humboldt Foundation and 
by the NSERC Canada.}

\end{document}